\documentclass[journal=jacsat,manuscript=article]{achemso}

\usepackage[version=3]{mhchem} 
\usepackage[table,xcdraw]{xcolor} 
\usepackage{graphicx} 

\author{Anjli Patel}
\affiliation[Stanford]
{Department of Chemical Engineering, Stanford University, Stanford, CA}
\author{Jens K. N\o rskov}
\affiliation[Stanford]
{Department of Chemical Engineering, Stanford University, Stanford, CA}
\alsoaffiliation[DTU]
{Department of Physics, Danish Technical University, Lyngby, Denmark}
\author{Kristin Persson}
\affiliation[LBL]
{Lawrence Berkeley National Laboratory, Berkeley, CA}
\affiliation[UCB]
{Department of Materials Science, University of California, Berkeley, CA}
\author{Joseph H. Montoya}
\email{joseph.montoya@tri.global}
\affiliation{Toyota Research Institute, Los Altos, CA}

\title[efficient Pourbaix]
  {Efficient Pourbaix diagrams of many-element compounds}

\abbreviations{IR,NMR,UV}

\begin{document}
Pourbaix diagrams are an invaluable tool for exploring the corrosion profiles of 
materials as a function of ambient pH and electrochemical potential\cite{Pourbaix1974ApplicationsPractice}. 
In recent years, high-throughput computational materials science efforts like those from the 
Materials Project \cite{Jain2013Commentary:Innovation, Ong2015ThePrinciples} have enabled more comprehensive 
Pourbaix diagrams to be constructed and disseminated from computational data
\cite{Persson2012PredictionStates, Singh2017ElectrochemicalMaterials}. These analyses have informed
a number of computational studies of materials for electrochemical applications,
aqueous electrocatalysis\cite{Rossmeisl2008ElectrocatalysisCalculations, Zhou2018RutileAcid, Marjolin2015ThermodynamicElectroreductions, Han2018ScreeningAcid} and 
photoelectrocatalysis\cite{Singh2019RobustDiscovery, E.2014NewCalculations, Yan2015Mninf2/infVinf2/infOinf7/inf:Splitting},
non-equilibrium crystallization \cite{Sun2019Non-equilibriumSolution, Wills2017GroupEnvironments}, 
and corrosion-resistant alloy design \cite{Ding2018ElectrochemicalStates, Huang2019ModelingDiagrams}.  In these, pourbaix analysis of multi-element systems is 
particularly valuable, as finding elusive materials like acid-stable oxygen evolution catalysts\cite{Zhou2018RutileAcid},
earth-abundant hydrogen evolution catalysts\cite{Hinnemann2005BiomimeticEvolution}, and selective \ce{CO2} reduction catalysts\cite{Torelli2016Nickel-Gallium-CatalyzedOverpotentials}
has and will likely continue to require exploration and optimization in multi-element spaces. 
However, Pourbaix analysis of phase stability on these resources have been 
limited to 3 or fewer elements, largely because computing the electrochemical phase 
stability of higher composition spaces has proven inefficient with existing methods. 

In this report, we provide details of a modified method for
Pourbaix diagram construction which enables diagrams to be constructed efficiently
in much higher compositional spaces, which enables phase stability analysis of similarly
complex individual materials. We demonstrate this functionality with an analysis of the
phase stability of a complex material for alkaline oxygen evolution (OER) and
highlight our implementation in the open-source pymatgen\cite{Ong2013PythonAnalysis}
code and on the Materials Project website (materialsproject.org).

The primary bottleneck in pymatgen's prior implementation of multi-element Pourbaix diagrams 
resides in their pre-processing iteration over potential combinations of compounds.  Essentially,
the current method, based on the thermodynamic formalisms outlined in ref\cite{Thompson2011PourbaixSystems} 
and ref\cite{Persson2012PredictionStates} is to iterate over all valid stoichiometric 
combinations of compounds in the chemical system
which satisfy the compositional constraint particular to a given Pourbaix diagram (e.g. Fe:Cr = 2:1).

In this scheme the scaling of Pourbaix diagram construction occurs with n choose m,
where n is the number of compounds included and m is the number of elements 
included. Since larger numbers of elements tend to produce more entries on queries of the database,
Pourbaix diagrams become prohibitively expensive after 3 elements. More explicitly, 4
or 5 element Pourbaix diagrams for the Ba-Sr-Co-Fe (present in BCSF\cite{Suntivich2011APrinciples}, an alkaline OER catalyst),
Al-Cu-Mn-Mg-Fe (present in some commercial Duralumin alloys) would require $\sim$$10^9$ and $\sim$$10^{11}$
evaluations of selected combinations of compounds from the pool of materials.

Considerable speedup is achieved by filtering for entries on the convex hull of the 
solid compositional phase diagram, which is at least partially motivated by physical 
reasoning that those materials should appear in the Pourbaix diagram absent any
ions. This process, however, is complicated by the variable chemical potential of \ce{H+} and
\ce{e-} on the Pourbaix diagram (but not on the compositional phase diagram) and the need to
add ionic species, which still results in poor combinatoric scaling. The process was also
further improved (e.g. in pymatgen) by virtue of it being easily parallelized, but this still
only renders a factor of N speedup when much larger factors are required for the higher-element 
spaces to be tractable.  In summary, with currently hardware, execution times for 5-element 
and higher diagrams are estimated to be on the order of years.

\begin{figure}
  \includegraphics[]{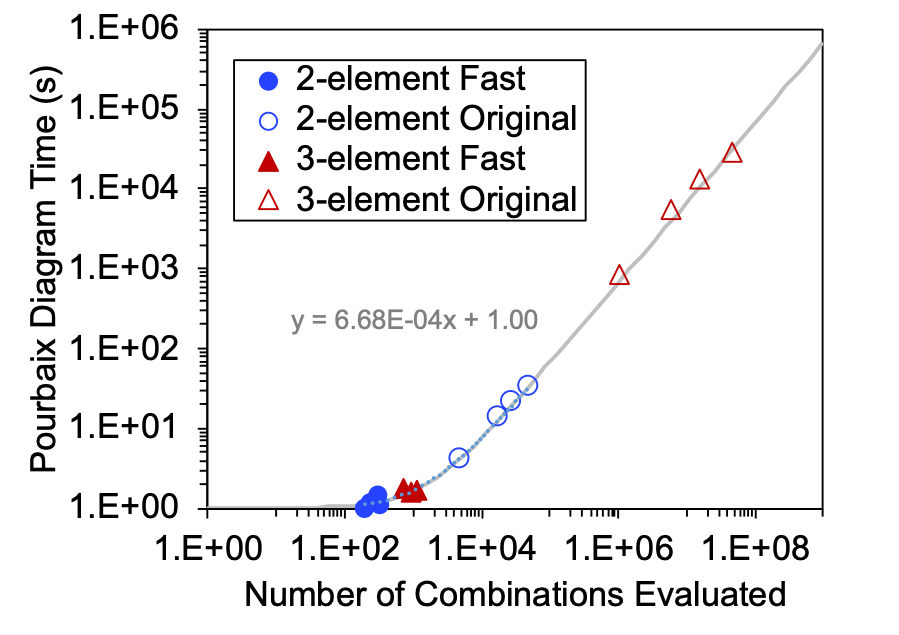}
  \caption{Relative scaling of current Pourbaix implementation, in which unfilled 
  points represent performance and iteration count of prior pymatgen implementation
  of pourbaix diagram construction.  Filled shapes represent performance of new 
  implementation.  Timing benchmarks for 4 and 5 element pourbaix diagrams are
  extrapolated based on scaling of 2 and 3-element performance with iteration 
  count}
  \label{fgr:performance}
\end{figure}

\begin{figure}
  \includegraphics[width=\textwidth]{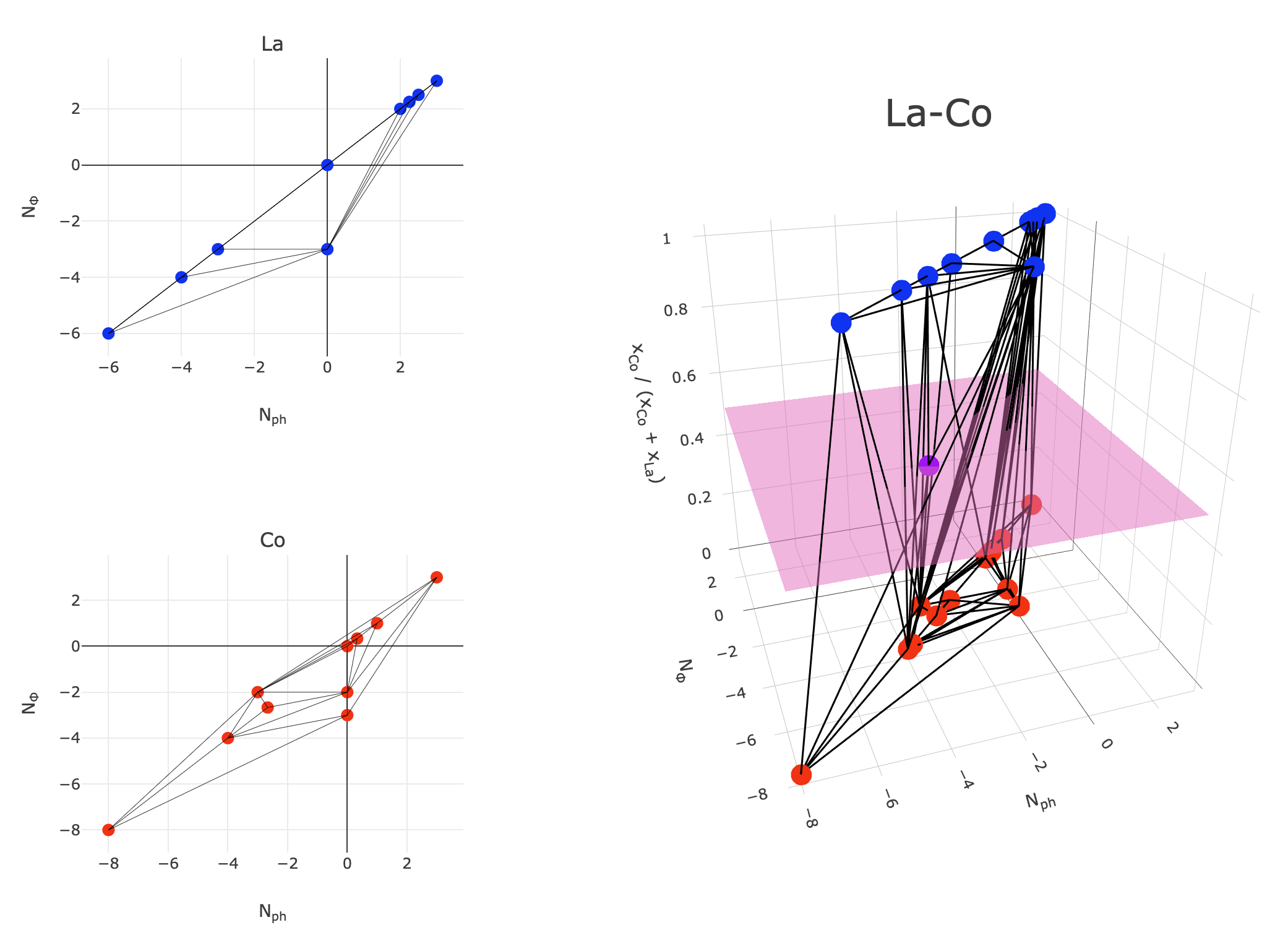}
  \caption{Convex hull projections for La, Co, and La-Co chemical systems 
  in $N_{pH}-N_\Phi$ and $N_{pH}-N_\Phi-x_{Co}$ space.
  The highlighted plane in the figure corresponding to the La-Co system
  represents the composition constraint at a fixed non-OH composition, 
  e. g. La:Co = 1:1 or $x_{Co} / (x_{La} + x_{Co}) = 0.5$.  Stable combinations
  of entries subject to this composition constraint may only be found in the 
  simplices of the multi-dimensional convex hull which intersect this 
  hyperplane.  Note that, in the 2-element case, mixed composition entries,
  for example the \ce{LaCoO3} shown in purple, appear in the interior
  of the simplicial complex.}
  \label{fgr:3dhull}
\end{figure}

To pre-filter the Pourbaix compounds that may appear on the hull, one can compute the convex 
hull in a similar manner as a pymatgen-implemented grand-canonical phase diagram, but in a space
which includes fractional coefficients on electrons and protons.  This essentially amounts to
a grand-canonical phase diagram in \ce{H+-e^--H2O-M1-M2-...-M$_n$}, for which valid stable (i. e. minimal free
energy of formation) compounds can be found by taking the convex hull in the space where $\mu_{\ce{H+}}$ and $\mu_{\ce{e-}}$ are treated
as free variables (i.e. points corresponding to their reference energies are not include in the convex 
hull point inputs).  For the purposes of finding stable
combinations of entries, a 4-D convex hull and its corresponding simplicies in 
$N_{pH}-N_{\Phi}-x_1-x_2 \ldots x_{n-1}$, where $N_{pH}$ and $N_{\Phi}$ are scaling factors 
for the Pourbaix energy\cite{Sun2019Non-equilibriumSolution} with respect to pH and applied 
potential, and $x_n$ are non-OH fractions of the non-OH elemental composition, are 
sufficient.  This hull and its corresponding simplices are illustrated for the La-Co 
Pourbaix system in Figure \ref{fgr:3dhull}.
Under the assumption of ideal mixing, decomposition products in this space correspond to 
simplices on the convex hull, meaning that valid Pourbaix decomposition products can be 
limited to those which appear in a given simplex.  The precise reduction in scaling will 
depend on the complexity of this hull, but it allows the combinatorial complexity to be 
isolated only to existing facets.   In practice, this offers a reduction in the
number of iterations by 2-3 orders of magnitude (see benchmarking in Figure 
\ref{fgr:performance}).  

We also note here that the determination of the Pourbaix regions in which the free energies
of the corresponding species is minimal are determined from a halfspace intersection of
2-dimensional planes corresponding to the pre-processed ``multi-entry" phases (as termed in pymatgen),
which differs from the grid-based methods implemented in ASE\cite{HjorthLarsen2017TheAtoms} 
and from ref\cite{Ding2018ElectrochemicalStates}.  However, our preprocessing might also be used 
to pre-filter compounds in a grid-based approach as these to reduce the iterative load at
each evaluated point in E-pH space.

To illustrate the power of the method, we benchmark the timing on the Pourbaix diagram derived from
the compound with the highest number of elements in the Materials Project database 
(\ce{Ba2NaTi2MnRe2Si8HO26F}, mp-1215061), which completes in 15-20 minutes.  This points 
to the added capability of featurizing the entire MP dataset with Pourbaix decomposition grids, 
which might make Pourbaix diagrams more amenable to emerging data-intensive prediction
methods such as machine and deep learning.

\begin{figure}
  \includegraphics[scale=0.5]{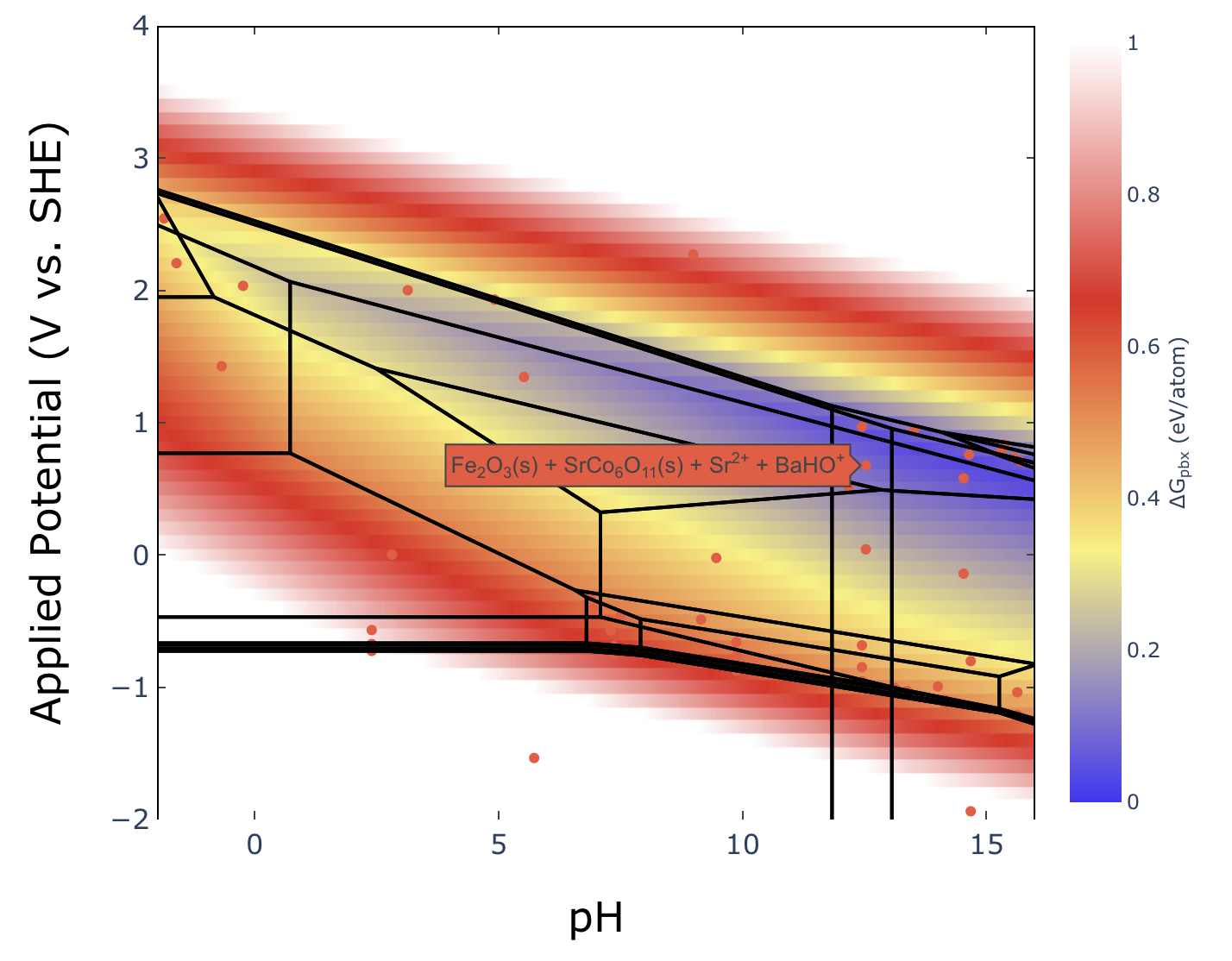}
  \includegraphics[scale=0.5]{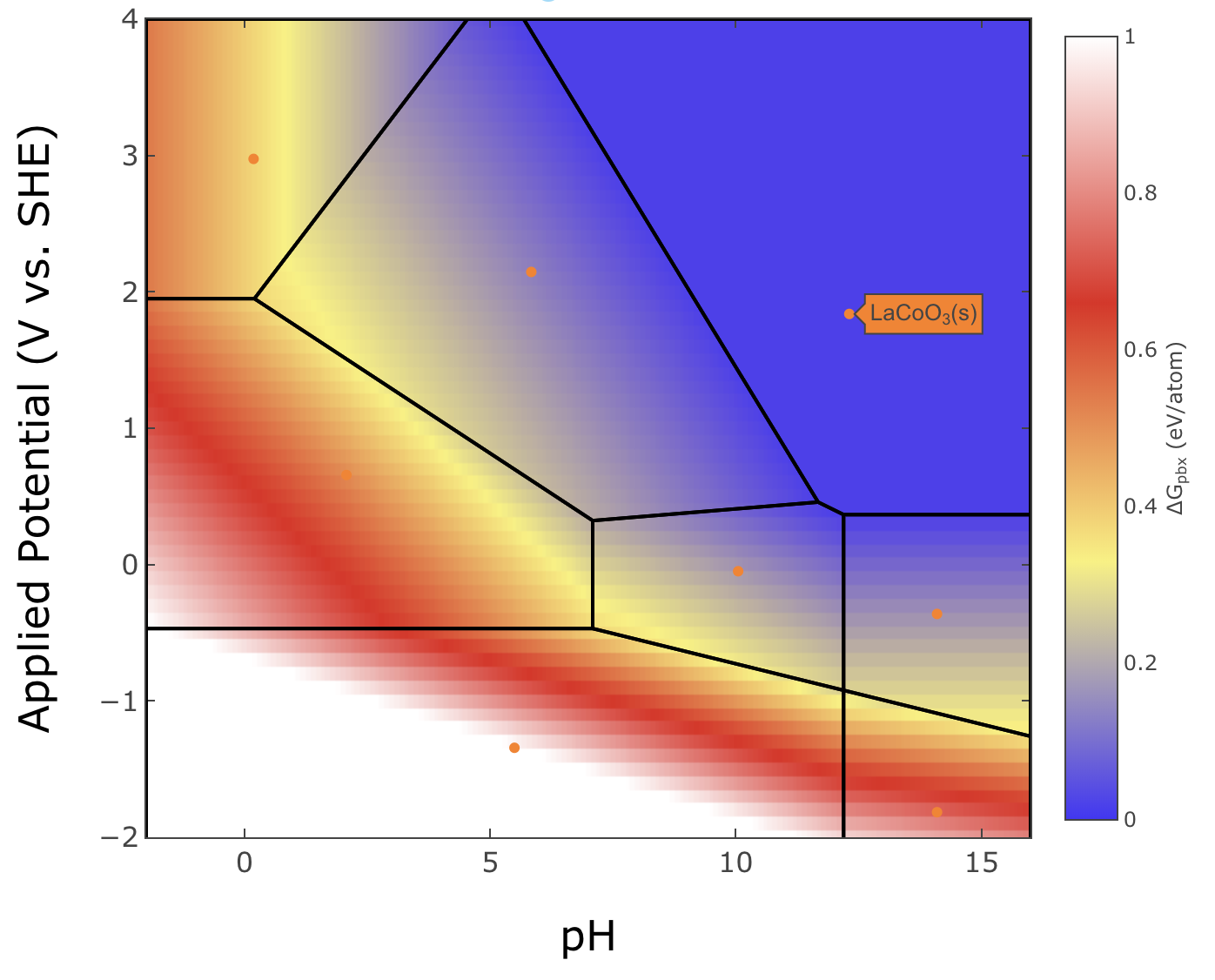}
  \caption{Pourbaix stability diagrams for (top) cubic perovskite \ce{BaSrCo7Fe7O24} (mp-1075935), 
  a model system for BSCF catalysts, and (bottom) \ce{LaCoO3}.  Decomposition energies in the alkaline OER 
  region are within the metastability window ($\sim$0.1 eV/atom), and the stable phases
  include ion-phase \ce{Sr^{2+}} and \ce{BaHO+}, indicating a modest driving force for cation leaching,
  whereas \ce{LaCoO3} retains its surface structure after alkaline OER 
  catalysis.\cite{Risch2013StructuralEXAFS,May2012InfluenceCatalysts}}
  \label{fgr:BSCF}
\end{figure}

To illustrate with another practical example, we include the heatmap corresponding to the
decomposition energy against the Pourbaix diagram (using the methods developed in
in ref\cite{Singh2017ElectrochemicalMaterials}) of 
the Ba-Sr-Co-Fe chemical system, a model system for the BSCF catalyst known for its 
high activity as an alkaline OER catalyst\cite{Suntivich2011APrinciples}.  
The Pourbaix diagram reveals that 
cation leaching is thermodynamically favorable with a modest driving force, suggesting 
that the material may not dissolve completely, but that the near-surface region 
may partially decompose in such a way that accounts for the experimental observations of surface
structure in post-OER characterization\cite{Risch2013StructuralEXAFS,May2012InfluenceCatalysts}.  
In Figure \ref{fgr:BSCF}, we present this in contrast to the \ce{LaCoO3} system which 
experiences no such corrosion, even after extensive cycling.  This materials's 
Pourbaix diagram suggests it has a large window of thermodynamic stability that is 
consistent with its experimentally observed stability.

To place this work in the appropriate context, we note that Pourbaix analysis may not tell 
the whole story of a given material's corrosion profile.  Kinetics also play a 
significant role in corrosion, notably in the role of the concentrations of various 
salts on the conductivity of the electrolyte and therefore electrochemical rates.
Additionally, the stability of a given passivation layer will frequently depend
on whether its inherent strain relative to the bulk material on which it forms
is energetically tolerable.  If not, as predicted by the Pilling-Bedworth ratio,
passivation layers will frequently flake or crack, which represents a corrosion-based
mode of material failure.\cite{Fromhold1976TheoryOxidation,Zhou1999FilmAlloys}  
Furthermore, surface stabilities differ from bulk 
stabilities, so the profile of nearest-surface region, which may be particularly relevant 
to the catalytic properties of a material, may exhibit subtle differences from the profile of
the bulk.  These differences notably manifest in the role of Pourbaix-dependent surface
coverage, which may influence reaction rates, particularly in alkaline OER.\cite{Lee2015AbNi,Hansen2008SurfaceDFT,Ulissi2016AutomatedLearning}.  
Finally, we note that the quality of a given pourbaix diagram will depend
on the quality of the thermodynamic data which is used to generate it. In the cases 
presented here, all of the input data is from DFT-computed formation energies, 
which have well-known and systematically correctable errors\cite{Jain2011FormationCalculations}.  
This dependence is complicated by the fact that the formation energies of ions in the
Materials Project scheme are computed relative to solids in order to allow for error cancellation
between ionic and solid formation energies\cite{Persson2012PredictionStates}.
However, pymatgen's software infrastructure is agnostic to the source of a given solid 
formation energy, and experimental formation energies may be used alone or in concert with
the computational data provided by the Materials Project API\cite{Ong2015ThePrinciples}.

In conclusion, we envision that this software functionality will have more 
general applications in the evaluation of corrosion resistance of complex alloys 
and of the stability of catalysts in high-throughput studies of water splitting 
and fuel cell reactions. As such, we have disseminated the implementation in the 
pymatgen.analysis.pourbaix\_diagram module of the pymatgen open-source software, 
enabling its use on the Materials Project website. Ultimately, it is our hope that 
efficient Pourbaix analysis of these complex compounds will enable new insights to be 
derived on materials which were previously intractable to analyze.

\begin{table}[]
\resizebox{\textwidth}{!}{%
\begin{tabular}{ccccc}
\hline
\rowcolor[HTML]{C0C0C0} 
\multicolumn{5}{|c|}{\cellcolor[HTML]{C0C0C0}{\color[HTML]{333333} \textbf{2-element}}}                                                                                                                                                                                                                                                                                                                       \\ \hline
\rowcolor[HTML]{C0C0C0} 
\multicolumn{1}{|c|}{\cellcolor[HTML]{C0C0C0}\textbf{System}} & \multicolumn{1}{c|}{\cellcolor[HTML]{C0C0C0}\textbf{Number of Fast Combos}} & \multicolumn{1}{c|}{\cellcolor[HTML]{C0C0C0}\textbf{Fast Pourbaix Time (s)}} & \multicolumn{1}{c|}{\cellcolor[HTML]{C0C0C0}\textbf{Number of Original Iterations}} & \multicolumn{1}{c|}{\cellcolor[HTML]{C0C0C0}\textbf{Original Pourbaix Time (s)}}           \\ \hline
\multicolumn{1}{|c|}{Bi-V}                                    & \multicolumn{1}{c|}{346}                                                    & \multicolumn{1}{c|}{1.1}                                                     & \multicolumn{1}{c|}{49974}                                                          & \multicolumn{1}{c|}{34.2}                                                                  \\ \hline
\multicolumn{1}{|c|}{Sr-Ir}                                   & \multicolumn{1}{c|}{203}                                                    & \multicolumn{1}{c|}{1.0}                                                     & \multicolumn{1}{c|}{4995}                                                           & \multicolumn{1}{c|}{4.0}                                                                   \\ \hline
\multicolumn{1}{|c|}{Cr-Fe}                                   & \multicolumn{1}{c|}{317}                                                    & \multicolumn{1}{c|}{1.4}                                                     & \multicolumn{1}{c|}{29463}                                                          & \multicolumn{1}{c|}{21.0}                                                                  \\ \hline
\multicolumn{1}{|c|}{Mo-S}                                    & \multicolumn{1}{c|}{247}                                                    & \multicolumn{1}{c|}{1.2}                                                     & \multicolumn{1}{c|}{18780}                                                          & \multicolumn{1}{c|}{13.6}                                                                  \\ \hline
                                                              &                                                                             &                                                                              &                                                                                     &                                                                                            \\ \hline
\rowcolor[HTML]{C0C0C0} 
\multicolumn{5}{|c|}{\cellcolor[HTML]{C0C0C0}{\color[HTML]{333333} \textbf{3-element}}}                                                                                                                                                                                                                                                                                                                       \\ \hline
\rowcolor[HTML]{C0C0C0} 
\multicolumn{1}{|c|}{\cellcolor[HTML]{C0C0C0}\textbf{System}} & \multicolumn{1}{c|}{\cellcolor[HTML]{C0C0C0}\textbf{Number of Fast Combos}} & \multicolumn{1}{c|}{\cellcolor[HTML]{C0C0C0}\textbf{Fast Pourbaix Time (s)}} & \multicolumn{1}{c|}{\cellcolor[HTML]{C0C0C0}\textbf{Number of Original Iterations}} & \multicolumn{1}{c|}{\cellcolor[HTML]{C0C0C0}\textbf{Original Pourbaix Time (s)}}           \\ \hline
\multicolumn{1}{|c|}{Fe-C-N}                                  & \multicolumn{1}{c|}{748}                                                    & \multicolumn{1}{c|}{1.8}                                                     & \multicolumn{1}{c|}{46191274}                                                       & \multicolumn{1}{c|}{29354.2}                                                               \\ \hline
\multicolumn{1}{|c|}{Zr-Ni-As}                                & \multicolumn{1}{c|}{1188}                                                   & \multicolumn{1}{c|}{1.8}                                                     & \multicolumn{1}{c|}{1078542}                                                        & \multicolumn{1}{c|}{838.9}                                                                 \\ \hline
\multicolumn{1}{|c|}{Ti-Al-Zn}                                & \multicolumn{1}{c|}{956}                                                    & \multicolumn{1}{c|}{1.6}                                                     & \multicolumn{1}{c|}{5979245}                                                        & \multicolumn{1}{c|}{5520.0}                                                                \\ \hline
\multicolumn{1}{|c|}{Ni-C-N}                                  & \multicolumn{1}{c|}{700}                                                    & \multicolumn{1}{c|}{1.7}                                                     & \multicolumn{1}{c|}{15879703}                                                       & \multicolumn{1}{c|}{13659.0}                                                               \\ \hline
                                                              &                                                                             &                                                                              &                                                                                     &                                                                                            \\ \hline
\rowcolor[HTML]{C0C0C0} 
\multicolumn{5}{|c|}{\cellcolor[HTML]{C0C0C0}{\color[HTML]{333333} \textbf{4-element}}}                                                                                                                                                                                                                                                                                                                       \\ \hline
\rowcolor[HTML]{C0C0C0} 
\multicolumn{1}{|c|}{\cellcolor[HTML]{C0C0C0}\textbf{System}} & \multicolumn{1}{c|}{\cellcolor[HTML]{C0C0C0}\textbf{Number of Fast Combos}} & \multicolumn{1}{c|}{\cellcolor[HTML]{C0C0C0}\textbf{Fast Pourbaix Time (s)}} & \multicolumn{1}{c|}{\cellcolor[HTML]{C0C0C0}\textbf{Number of Original Iterations}} & \multicolumn{1}{c|}{\cellcolor[HTML]{C0C0C0}\textbf{Projected Original Pourbaix Time (s)}} \\ \hline
\multicolumn{1}{|c|}{Ba-Sr-Fe-Co}                             & \multicolumn{1}{c|}{4409}                                                   & \multicolumn{1}{c|}{4.4}                                                     & \multicolumn{1}{c|}{4083216060}                                                     & \multicolumn{1}{c|}{3083969}                                                               \\ \hline
                                                              &                                                                             &                                                                              &                                                                                     &                                                                                            \\ \hline
\rowcolor[HTML]{C0C0C0} 
\multicolumn{5}{|c|}{\cellcolor[HTML]{C0C0C0}{\color[HTML]{333333} \textbf{5-element}}}                                                                                                                                                                                                                                                                                                                       \\ \hline
\rowcolor[HTML]{C0C0C0} 
\multicolumn{1}{|c|}{\cellcolor[HTML]{C0C0C0}\textbf{System}} & \multicolumn{1}{c|}{\cellcolor[HTML]{C0C0C0}\textbf{Number of Fast Combos}} & \multicolumn{1}{c|}{\cellcolor[HTML]{C0C0C0}\textbf{Fast Pourbaix Time (s)}} & \multicolumn{1}{c|}{\cellcolor[HTML]{C0C0C0}\textbf{Number of Original Iterations}} & \multicolumn{1}{c|}{\cellcolor[HTML]{C0C0C0}\textbf{Projected Original Pourbaix Time (s)}} \\ \hline
\multicolumn{1}{|c|}{Al-Mn-Fe-Cu-Mg}                          & \multicolumn{1}{c|}{26310}                                                  & \multicolumn{1}{c|}{29.8}                                                    & \multicolumn{1}{c|}{7.04625E+12}                                                    & \multicolumn{1}{c|}{5321889378}                                                            \\ \hline
\end{tabular}
}
\caption{Performance Metrics for Fast Algorithm}
    \label{tab:performancetable}
\end{table}

\bibliography{references}

\end{document}